\documentclass[a4paper,conference]{IEEEtran}
\IEEEoverridecommandlockouts
\usepackage{subcaption}
\usepackage{fancyhdr,setspace}
\usepackage{comment}
\usepackage{caption}
\usepackage{todonotes}
\usepackage{bm}
\renewcommand{\footnoterule}{}

\usepackage[compact]{titlesec} 
\usepackage{cite}
\usepackage{xcolor}
\usepackage{blindtext, graphicx, tabularx, multicol, lipsum, subfig,eqnarray,amsmath,chemformula,multirow,threeparttable}
\usepackage{enumitem}
\usepackage[utf8]{inputenc}
\usepackage{amssymb,amsmath}

\usepackage{float}

\usepackage[boxed]{algorithm2e}
\usepackage[noend]{algpseudocode}
\usepackage[compact]{titlesec}
\algnewcommand{\IfThenElse}[3]{
  \State \algorithmicif\ #1\ \algorithmicthen\ #2\ \algorithmicelse\ #3}
\makeatletter
\newcommand{\removelatexerror}{\let\@latex@error\@gobble}
\makeatother
\makeatletter
\renewcommand\footnoterule{%
  \kern-3\p@
  \hrule\@width\columnwidth
  \kern2.6\p@}
  \makeatother
\usepackage{letltxmacro}

\usepackage{amsthm}

\usepackage{amsmath}

\SetKwInput{KwInput}{Input}                
\SetKwInput{KwOutput}{Output}  

\usepackage{tikz}
\usepackage{capt-of}
\usepackage{amssymb} \usepackage{booktabs} 
\usepackage{pifont}
\newcommand{\xmark}{\ding{55}}%

\usepackage{diagbox}
\usepackage{xcolor,etoolbox}
\usepackage{dblfloatfix}

\let\mybibitem\bibitem
\renewcommand{\bibitem}[1]{%
\ifstrequal{#1}{edgeTPU}{\color{black}\mybibitem{#1}}
{\ifstrequal{#1}{xyz}{\color{blue}\mybibitem{#1}}
{\color{black}\mybibitem{#1}}}%
}

\ifCLASSINFOpdf
 
\else

\fi

\usepackage{comment}

\begin{document}

%

\title{FLARE: \underline{F}ault Attack \underline{L}everaging \underline{A}ddress \underline{R}econfiguration \underline{E}xploits in Multi-Tenant FPGAs$^*$\thanks{$^*$The work of J. Chaudhuri and K. Chakrabarty was supported in part by the National Science Foundation under grant no. CNS-2011561. The work of Hassan Nassar was supported in part by the German Federal Ministry of Education and Research (BMBF) through grant 01IS23066 as part of the Software Campus Project ``HE-Trust'' and in part by the ``Helmholtz Pilot Program for Core Informatics (kikit)'' at KIT. The work of Mehdi Tahoori was supported by German Research Foundation (DFG) projects SAUBER and SecFShare.}}

\author{\IEEEauthorblockN{Jayeeta Chaudhuri$^\dagger{}$, Hassan Nassar$^{\ddagger}{}$, Dennis R.E. Gnad$^{\ddagger}{}$, Jörg Henkel$^{\ddagger}{}$, \\
Mehdi B. Tahoori$^{\ddagger}{}$,  and Krishnendu Chakrabarty$^\dagger{}$}
\IEEEauthorblockA{$^\dagger{}$School of Electrical, Computer, and Energy Engineering, Arizona State University, Tempe, AZ, USA \\
$^\ddagger{}$Institute of Computer Engineering, Karlsruhe Institute of Technology (KIT), Karlsruhe, Germany
\\
}
\\[-5.0ex]
}

\maketitle

\begin{abstract}

Modern FPGAs are increasingly supporting multi-tenancy to enable dynamic reconfiguration of user modules. While multi-tenant FPGAs improve utilization and flexibility, this paradigm introduces critical security threats. In this paper, we present \textbf{FLARE}, a fault attack that exploits vulnerabilities in the partial reconfiguration process, specifically while a user bitstream is being uploaded to the FPGA by a reconfiguration manager. Unlike traditional fault attacks that operate during module runtime, FLARE injects faults in the bitstream during its reconfiguration, altering the configuration address and redirecting it to unintended partial reconfigurable regions (PRRs). This enables the overwriting of pre-configured co-tenant modules, disrupting their functionality. FLARE leverages power-wasters that activate briefly during the reconfiguration process, making the attack stealthy and more challenging to detect with existing countermeasures. Experimental results on a Xilinx Pynq FPGA demonstrate the effectiveness of FLARE in compromising multiple user bitstreams during the reconfiguration process.
\end{abstract}


\thispagestyle{plain}
%
\IEEEpeerreviewmaketitle

\section{Introduction}

Field Programmable Gate Arrays (FPGAs) are increasingly being used as customized accelerator platforms to support high-performance computing and accelerating compute-intensive workloads. Major cloud service providers (CSPs) such as Amazon, Alibaba Cloud, and Microsoft offer FPGA-accelerated instances, which allow users to deploy their custom modules with high flexibility and efficiency, particularly for artificial intelligence and machine learning applications~\cite{ref_17_amazon, firestone2018azure}. With their scalability across several compute-intensive tasks, FPGAs are being increasingly adopted for virtualization and multi-tenancy applications. The multi-tenant approach allows CSPs to maximize resource utilization by allowing users access to simultaneously reconfigure their applications on separate modules of the same FPGA.  

However, the multi-tenancy approach introduces several security threats, particularly physical and remote attacks. Physical attacks occur when an attacker gains direct access to the FPGA system, enabling them to measure power consumption, gather timing information, or measure electrical current~\cite{moradi2011vulnerability, standaert2004power, standaert2006updates}. However, emerging threats demonstrate that fault attacks can be activated remotely via untrusted users using on-chip sensors that impact the functionality of co-tenant modules, without requiring physical access to the FPGA~\cite{trimberger2017security}. In~\cite{8056840}, ring oscillator (RO)-based circuits have been used to create excessive voltage fluctuations, that lead to a permanent crash of the FPGA. Building on this, \cite{FPGAhammer, 7809042, 8844478} showed that by repeatedly toggling a set of ROs, attackers can induce fault attacks on an AES module. In~\cite{chaudhuri2024hackingfabrictargetingpartial}, certain toggling frequencies of an RO grid were shown to induce fault attacks during partial reconfiguration on an FPGA. Faults are injected into the bitstream by targeting the reconfiguration manager (RM), which manages the upload and configuration of bitstreams in the partial reconfigurable regions (PRRs) of the FPGA. Unlike~\cite{8056840, FPGAhammer, 7809042, 8844478}, these faults persist even after the attack ceases, resulting in computational errors after FPGA deployment.

Prior runtime fault attacks~\cite{8056840, FPGAhammer, 7809042, 8844478} require extended attack durations, as power-wasters need to remain activated throughout the operation of the tenant modules. This requirement increases the likelihood of attack detection by voltage-monitoring schemes~\cite{9643485, 10.1145/3451236}. In~\cite{chaudhuri2024hackingfabrictargetingpartial}, the attack duration is reduced by activating power-wasters only during partial reconfiguration. However,~\cite{chaudhuri2024hackingfabrictargetingpartial} is successful only when Cyclic Redundancy Check (CRC) of a bitstream is disabled, which makes the attack more detectable. Moreover, all the aforementioned attacks are limited to disrupting only a single tenant module. 

In this paper, we introduce FLARE, a novel fault attack that adversely impacts multiple tenants by altering the reconfiguration address of a user bitstream rather than the bitstream data. To achieve this, we employ malicious power-wasters that induce faults in the bitstream while it is being configured onto the FPGA through the RM. These faults alter the `select' part of the bitstream, which specifies the configuration address that determines the exact PRR where the bitstream is to be loaded \cite{9643485}. By injecting faults into the configuration address, the bitstream is incorrectly redirected, resulting in the overwriting of PRRs of other co-tenants. This leads to erroneous computation and potential Denial-of-Service (DoS) for the user bitstream. The `select' part of the bitstream is much smaller than the full bitstream, consisting of only a few configuration words. By precisely targeting fault injection in the `select' part, the attack duration is significantly minimized, making it \textit{stealthy} and harder to detect by the schemes described in \cite{10.1145/3451236, 9643485}. In our paper, a stealthy attack is defined as the attacker's ability to evade detection by countermeasures that rely on monitoring extended fault injection activity.

\begin{figure}

\includegraphics[width=0.48\textwidth]{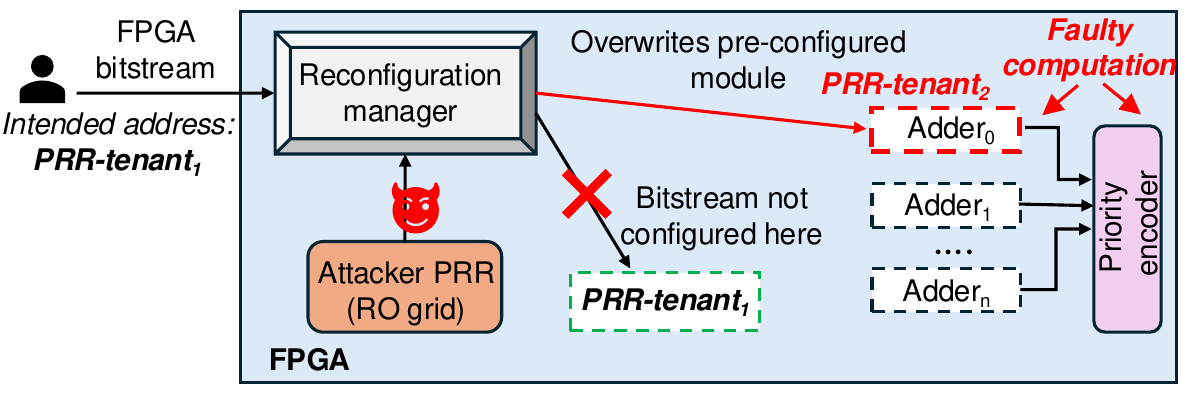}

\caption{Proposed attack setup of FLARE (Adder$_0$: benign module configured in address \textit{PRR-tenant$_2$}. Due to fault-injection, an user bitstream is redirected to \textit{PRR-tenant$_2$} instead of its intended address \textit{PRR-tenant$_1$}).}
\label{flare}
\vspace{-0.6cm}

\end{figure}

We demonstrate the effectiveness of FLARE using benign design modules that are configured within a multi-tenant FPGA setup. Fig. \ref{flare} illustrates the proposed fault attack setup.

The key contributions of this paper are as follows.
\begin{itemize}[leftmargin=*,topsep=0pt]
\item We determine the appropriate configuration parameters of several power-wasters to induce a successful fault attack.
    \item We demonstrate the first address manipulation fault attack that targets the process of FPGA partial reconfiguration.
    \item We show that targeted fault injections in the configuration address section of the bitstream can disrupt the functionality of other modules sharing the FPGA fabric.
    \item We evaluate FLARE on several modules, including an AES cryptography implementation and adder instances.
    \item We implement FLARE on hardware, targeting the Xilinx Pynq-Z1 FPGA board.
    
\end{itemize}

The remainder of the paper is organized as follows. Section II provides an overview of prior attacks on multi-tenant FPGAs and establishes the motivation of our attack. The threat model is described in Section III. Section IV presents the attack framework and the experimental setup. Section V provides experimental results. Finally, Section VI concludes the paper.
\vspace{-0.5cm}

\section{Related Prior Work and Motivation}
\vspace{-0.1cm}
\subsection{Prior Fault Attacks on Multi-Tenant FPGAs}
\vspace{-0.1cm}

In FPGAs, PRRs are designed to maintain logical isolation to ensure secure operation for co-tenant modules \cite{4223233}.
However, recent work explores the vulnerabilities of multi-tenant FPGAs to remotely induced fault attacks; these attacks do not require physical access to the FPGA. An adversary can implement power-wasters on the FPGA to exploit the power distribution network (PDN), which supplies power to all the tenant modules that are configured on the FPGA. This results in rapid voltage fluctuations on the shared PDN.

In \cite{8056840}, authors implemented ROs for generating abrupt voltage drops and showed their effectiveness in DoS and crashing of the FPGA. Extending upon this approach, \cite{FPGAhammer} explored the potential of ROs to induce voltage drop-based fault-injections in an AES module. By selecting the appropriate RO toggling frequency, faults can be injected in AES, resulting in incorrect ciphertext generation or key recovery. \cite{Sugawara2019OscillatorCentre, provelengios2020power, Krautter2019MitigatingCloud} explored non-combinational ROs and glitch amplification circuits that evade the Amazon Web Services design rule check (DRC).
In \cite{8844478}, a runtime fault attack was demonstrated that did not require combinational loops or glitch-inducing designs. Instead, the attack exploited short circuits within the FPGA to force simultaneous writes to the same memory address in a dual-port RAM. These memory collisions caused permanent alterations in the bitstream configuration. In contrast to the aforementioned attacks that are executed at runtime, \cite{chaudhuri2024hackingfabrictargetingpartial} explores a fault attack that is carried out during the process of bitstream reconfiguration. In this approach, faults are injected directly into the bitstream, which persist even after the bitstream is configured onto the FPGA.

\vspace{-0.1cm}
\subsection{Fault Attack Countermeasures}
\vspace{-0.1cm}
Several countermeasures in the literature aim to detect fault attacks that employ power-wasters; these countermeasures are focused on identifying runtime fault attacks. In \cite{9643485}, fault attacks are detected using voltage drop-based sensors. This technique aims at disabling the interconnects of the attacker module, blocking it from FPGA configuration. Another fault-monitoring scheme, presented in \cite{10.1145/3451236}, disables the clock signal of the fault-impacted tenant to halt the activity of power-wasters. 
These methods rely on power-wasters being continuously activated for a significantly long duration to detect fault attacks. Note that all prior runtime fault attacks \cite{8056840, FPGAhammer, 7809042, 8844478} persist throughout the victim module's operation. Such prolonged attack durations increase the likelihood of detection by the techniques presented in \cite{9643485, 10.1145/3451236}. 

In contrast, FLARE is stealthier and faster, with a higher likelihood of evading known countermeasures. FLARE minimizes the attack duration by precisely activating the power-wasters only when the bitstream configuration address is being loaded into the RM. Unlike \cite{chaudhuri2024hackingfabrictargetingpartial}, which is detected by CRC, FLARE evades CRC detection successfully. 
Table \ref{comp} presents a qualitative comparison of FLARE with prior fault attacks.



\begin{table}[t]
\centering
\fontsize{7.3}{7.3}\selectfont

\caption{Comparison of FLARE with prior fault attacks on multi-tenant FPGAs (Note: Few $\mu$s (ms) refers to a duration of a few microseconds (milliseconds)).}

    \label{compare}
    \begin{tabular}{|c| c| c| c| c|c|} 
    \hline
{Attack} &Target & Loop-free  &Attack &Address&Tenant \\
method&module&oscillator &duration&altered?&impacted \\
&&evaluated?&&& \\
\hline
\cite{8844478}&RAM&\xmark&Continuous&\xmark&Single \\

\cite{FPGAhammer}&AES&\xmark&Continuous&\xmark&Single \\

\cite{8056840}&FPGA&\xmark&Continuous&\xmark&Single \\

\cite{7809042}&Bitstream&\xmark&Continuous&\xmark&Single\\

\cite{chaudhuri2024hackingfabrictargetingpartial}&Bitstream&\xmark&Few ms&\xmark&Single \\

\cite{provelengios2020power}&RSA&\xmark&Continuous&\xmark&Single \\

\cite{Krautter2019MitigatingCloud} &FPGA&\checkmark&Continuous&\xmark&Single \\
\multirow{2}{*}{\textbf{FLARE}} &Bitstream, &\multirow{2}{*}{\checkmark}&\multirow{2}{*}{Few $\mu$s}&\multirow{2}{*}{\checkmark}&\multirow{2}{*}{Multiple} \\
&FPGA PRR(s)&&&& \\

\hline
\end{tabular}

\vspace{-0.4cm}

\label{comp}
\end{table}

\section{Threat Model}

We consider a multi-tenant FPGA setup, i.e., multiple users (tenants) deploy their modules independently on different regions of the FPGA. In this scenario, an attacker is a malicious third-party user who reconfigures one of the PRRs of the FPGA with a malicious power-waster. The goal of the attacker is to manipulate the configuration address of the user bitstream while it is being loaded onto the RM. The RM, which is configured on the same FPGA, loads the bitstream in its memory and subsequently configures it onto the designated PRRs of the FPGA. However, the RM is not designed with security as a focus, making it vulnerable to exploitation. 

\textbf{Attacker Knowledge:} The RM includes a status register that can be read by  applications and tenants without requiring access privileges \cite{7946114}. An attacker can monitor this status  to detect when a bitstream is being loaded into the RM, including the critical timing of loading the configuration address.

\textbf{Attacker Capability:} The attacker activates the power-wasters when the  configuration address of the bitstream is being loaded into the RM. Upon activation of the power-wasters, there are significant voltage fluctuations in the PDN, which induces faults in the bitstream.

\textbf{Attack Outcome:} The injected faults in the bitstream alter its configuration address, redirecting it to the wrong PRR. 

Note that the RM is capable of loading both encrypted and unencrypted bitstreams. However, bitstreams are typically decrypted before FPGA configuration \cite{ref_ultra}. Therefore, in our experiments, we specifically focus on decrypted bitstreams.

\section{Fault Attack Targeting Reconfiguration Address Manipulation}
\vspace{-0.1cm}
In this section, we provide a detailed analysis of FLARE, which targets configuration address manipulation. First, we implement several power-wasters that have been shown to cause voltage-based fault attacks and DoS \cite{8056840, FPGAhammer, Krautter2019MitigatingCloud, 9810438}. We evaluate the appropriate configuration parameters of these power-wasters for triggering a successful fault attack on the FPGA. Next, we describe how activation of these power-wasters during the partial reconfiguration process of bitstreams can adversely impact co-tenant modules on the FPGA. Finally, we provide a detailed description of the bitstream reconfiguration process and the attack setup on the FPGA.
\vspace{-0.1cm}
 \subsection{Fault Injection Using Power-Wasters }
\vspace{-0.1cm}
 As demonstrated in \cite{8056840, chaudhuri2024hackingfabrictargetingpartial}, ROs can be implemented as malicious power-wasters, causing high voltage fluctuations when triggered at a specific toggling frequency. Alternatively, loop-free self-clocked ROs  bypass DRC and are capable of generating clock glitches and severe voltage fluctuations on the PDN \cite{Sugawara2019OscillatorCentre}. Based on these observations, we evaluate both combinational and loop-free ROs for injecting faults into bitstreams. To induce malicious fault injections in bitstreams during partial reconfiguration, it is crucial to determine appropriate fault-injection parameters for the RO grid: (1) Number of ROs, (2) toggling frequency, and (3) duty cycle. We choose the upper bound of the number of RO and self-clocked RO instances to be 16000 and the toggling frequency range as $10^5-10^6$ Hz to be consistent with the experimental setup in prior work \cite{chaudhuri2024hackingfabrictargetingpartial, gnad2020remote}. These power-wasters are deployed on the FPGA to launch fault attacks on user bitstreams while they are being loaded into the RM for partial reconfiguration.
\vspace{-0.1cm}
\begin{figure}
\centering
\includegraphics[width=0.32\textwidth]{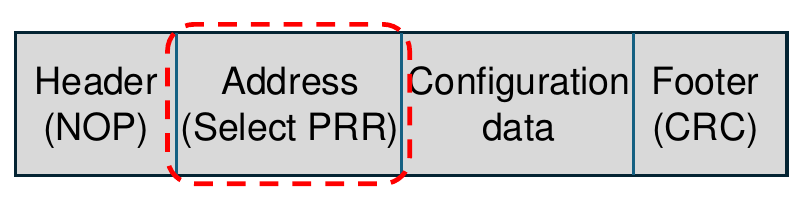}

\caption{Structure of a Xilinx partial bitstream (the target region for attack is highlighted in red).}
\label{structure}
\vspace{-0.5cm}
\end{figure}
 \subsection{Proposed Fault Attack Methodology}
\vspace{-0.1cm}
 Prior fault attacks \cite{FPGAhammer, 8056840} require the power-wasters to be activated continuously throughout the runtime of user modules to ensure a successful  fault attack. This significantly increases the attack \textit{exposure window}, i.e., the duration for which the attack needs to be active, making it more susceptible to detection. Unlike previous approaches, FLARE is precisely timed to be activated only when the configuration address of the bitstream is being loaded to the RM and remains inactive otherwise. This targeted activation makes the attack stealthy and minimizes the likelihood of detection \cite{9643485, 10.1145/3451236}. By injecting faults in `select' part of the bitstream, FLARE manipulates the configuration address, redirecting the bitstream to incorrect PRR(s), which we refer to as \textit{victim} PRRs.
 
\textbf{Injecting Faults to the Reconfiguration Address:} To understand how FLARE operates, it is important to examine the structure of an FPGA bitstream. Fig. \ref{structure} shows the bitstream structure of a Xilinx 7-series FPGA. The bitstream is partitioned into several structured frames \cite{pham2017bitman}.  The first section is the synchronization header, which initializes the bitstream. The header is followed by configuration address frames, which allocate a fixed set of bits namely `select' \cite{9643485}. Finally, the footer includes CRC values. The `select' part identifies the target PRRs where the bitstream is intended to be uploaded. The precise timing of FLARE is based on the fixed and well-defined structure of the Xilinx FPGA bitstream \cite{9643485}. The configuration process of a bitstream on the FPGA operates at a specific clock frequency. Using this frequency and information about the number of frames prior to the `select' frames, an attacker can estimate the time window during which the `select' part will be configured. In this duration, the attacker activates the RO grid to inject faults in the bitstream. 
 
\textbf{Analysis of Attack Impact:} We emphasize that while earlier attacks primarily focused on single-tenant vulnerabilities or localized faults within a specific module, FLARE exposes a broader vulnerability by targeting multiple modules co-located on the FPGA, a threat that was not explored in prior fault attacks \cite{chaudhuri2024hackingfabrictargetingpartial, FPGAhammer, krautter2021remote, provelengios2020power}. Importantly, we evaluate the effectiveness of the attack after deactivating the power-wasters. At this point, the faulty bitstreams are already loaded into the FPGA, and any errors in module functionality are observed post-reconfiguration. The objectives of FLARE are as follows:

\begin{itemize}[leftmargin=*,topsep=0pt]
    \item \textit{Denial-of-service}:  By manipulating the configuration address of the user bitstream, the bitstream is redirected to incorrect PRRs. This results in an incomplete configuration of the intended module, rendering it non-functional. 
    \item \textit{Faulty computation in co-tenant module(s)}: When a bitstream is redirected to a wrong configuration address, it may overwrite the existing logic of a co-tenant module. This leads to erroneous computations and incorrect functionality of the module. For instance, arithmetic modules used for data processing may generate incorrect results, potentially compromising the integrity and functionality of user modules. Note that once the faults are injected in the victim PRRs, they persist until the FPGA is fully reconfigured. 
\end{itemize}

\vspace{-0.2cm}

\subsection{Attack Setup}
\vspace{-0.1cm}
\begin{figure}[t]
\includegraphics[width=1\linewidth]{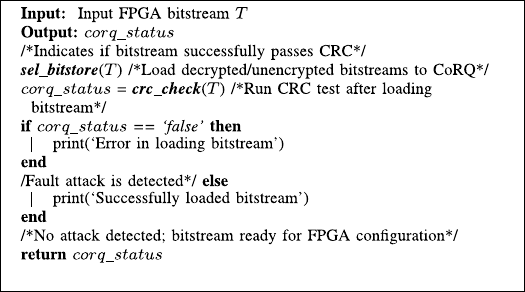}

\caption{Pseudocode for partial reconfiguration of bitstreams on the FPGA via CoRQ.}\label{alg}

\vspace{-0.6cm}
\end{figure}

\textbf{Dynamic Partial Reconfiguration Using CoRQ:} We utilize an open source runtime RM, namely Command-based Reconfiguration Queue (CoRQ), for loading and dynamically reconfiguring bitstreams on the FPGA \cite{7946114}. CoRQ can store multiple partial bitstreams corresponding to different modules, and reconfigure them onto specific PRRs of the FPGA based on their configuration addresses. The steps for dynamic reconfiguration of a bitstream are illustrated in Fig. \ref{alg}. First, the FPGA bitstreams are loaded into the CoRQ memory (marked as \textbf{\textit{sel\_bitstore}}). Next, they are passed through CRC evaluation (marked as \textbf{\textit{crc\_check}}). If CRC passes, the bitstreams are successfully configured on the FPGA, else they are blocked from reconfiguration.
Vivado allows users to enable or disable the CRC of bitstreams based on their requirements. In scenarios where the bitstream is encrypted, the CRC is disabled using the constraint \textit{BITSTREAM.GENERAL.CRC Disable} \cite{ref_ultra}. To assess the impact of FLARE on CRC, we deliberately enable CRC during bitstream generation in all our experiments.

CoRQ shares its status as a register that can be read by the processing system, enabling real-time guarantees that can be exploited by attackers. By monitoring when CoRQ starts to be busy, we can precisely determine when the configuration address  begins to be uploaded (explained in Section IV.B).

Note that while our experiments target a specific RM, the attack methodology is generic and applicable to various RMs in FPGAs as the reconfiguration always has the same flow.

\textbf{Implementation of Victim Modules:} We evaluate FLARE on two design modules: (1) Adders, which are commonly used in ALUs and signal processing, and (2) AES, which is needed in communication and cryptography operations \cite{krautter2021remote}.  To create a realistic attack scenario, we configure a majority of the lookup tables (LUTs) on the FPGA with these modules \cite{luo2020stealthy}. This reflects real-world multi-tenant FPGAs, where shared modules boost throughput and utilization \cite{mbongue2020architecture}.
\vspace{-0.1cm}
\subsubsection{Case Study 1: Adder}
\vspace{-0.1cm}
In our first case study, we evaluate combinational adders as their outputs are not impacted by timing violations commonly associated with fault attacks \cite{9643485}, allowing us to specifically focus on the impact of address manipulation. To identify which adders are affected by the fault-injected input bitstream, we implement two adder clusters, referred to as adder cluster \#1 and adder cluster \#2, with each cluster containing $n$ adder modules. Through experimental analysis, we observe that initializing adders separately rather than grouping them into clusters is not feasible, as it results in inefficient resource utilization.  To ensure maximum LUT usage, we set $n=500$; $n>500$ causes placement error during Vivado implementation. Each adder module computes the sum of fixed inputs. In an address manipulation-based fault attack, the input bitstream is redirected to the adder PRRs, leading to erroneous computations in the adder modules. To localize the faulty adders, we generate signals $flag_1^i$ and $flag_2^i $, $1 \leq i \leq n$, corresponding to each adder module $i$ in adder clusters \#1 and \#2, respectively. A value of `0' for $flag_1^i$ ($flag_2^i$) indicates that adder $i$ in adder cluster \#1 (adder cluster \#2) produces the correct output, while a value of `1' indicates a faulty output.

Monitoring the $flag_1$ and $flag_2$ signals directly is not practical due to their large widths of 500 bits each. To address this, we encode $flag_1$  and $flag_2$ using priority encoder modules $p_1$ and $p_2$, respectively; this enables us to localize the faulty adder modules within the respective adder clusters. The adder clusters and priority encoders contribute to 21.2\% of LUT utilization. Finally, we read the data from $flt_{sig}$, which encodes the information about the faulty adders. Specifically, the lower 10 bits of $flt_{sig}$, denoted as $flt_{sig}[9:0]$, are used by $p_1$ to indicate the faulty adder from adder cluster \#1. Similarly, the next 10 bits, $flt_{sig}[19:10]$, are used by  $p_2$ to denote the faulty adder from adder cluster \#2.

\vspace{-0.1cm}

\subsubsection{Case Study 2: AES}
\vspace{-0.1cm}
In our second case study, we evaluate FLARE on an AES-128 implementation. AES encryption and decryption involve the following operations -- \textit{sub\_bytes}, \textit{mix\_columns}, \textit{shift\_rows}, and \textit{key\_expansion}. Given the substantial overhead of each AES module in terms of the number of configured LUTs on the FPGA (each AES module utilizes 10223 LUTs), we implement only two such AES instances, namely AES \#1 and AES \#2, which contribute to 38.6\% of the FPGA LUT utilization. For our experiments, we configure both the AES modules with the same 128-bit plaintext and 128-bit key. A fault attack is successful when the fault-injected bitstream is redirected to atleast one of the AES modules, resulting in computational errors. We compare the outputs of the AES modules against the predetermined ciphertext. If the ciphertext of AES \#1 (AES \#2) matches the expected value, we set $flag_1$ ($flag_2$) as 0, else we set it as 1. The fault detection signals
$flag_1$ and $flag_2$ are encoded in the least significant bits $flt_{sig}[0]$ and $flt_{sig}[1]$, respectively. The value of $flt_{sig}$ determines one of the following:
\begin{itemize}[leftmargin=*,topsep=0pt]
    \item $flt_{sig}=0$: No fault attack is detected in the AES modules.
    \item $flt_{sig}=1$: A fault attack results in incorrect reconfiguration of a bitstream in the PRR of AES \#1. 
    \item $flt_{sig}=2$: A fault attack results in redirection of the bitstream to AES \#2. 
    \item  $flt_{sig}=3$: Both AES \#1 and AES \#2 are impacted. 
\end{itemize}
\vspace{-0.1cm}

\section{Experimental Results}
\subsection{Experimental Setup}
\vspace{-0.1cm}
\begin{table}[t]
\centering
\caption{Resource utilization of different fault attack setups on Pynq-Z1 FPGA (SC: Self-clocked RO).}
\fontsize{8.2}{8.2}\selectfont

\begin{tabular}{|c |c |c|c|c|c|} 
\cline{1-6}
{Module}&No. of &\multicolumn{2}{c|}{LUT configured by  }&\multicolumn{2}{c|}{Overall LUT } \\ 

&instances&\multicolumn{2}{c|}{power-wasters (\%)}& \multicolumn{2}{c|}{utilization (\%)}\\
\cline{3-6}
&&RO&SC &RO&SC  \\
\hline
Adder&1000&29.6&28.2&64.6&76.9 \\
\hline
AES&2&22&26&76.7 &78.7\\
\hline

\end{tabular}
 \vspace{-0.2cm}
 \label{lut}

\end{table}

We implement FLARE on the Pynq-Z1 FPGA, which consists of 53,200 LUTs. The power-wasters, AES, and adder modules are implemented in Verilog. Note that while our evaluation focuses on RO and self-clocked RO circuits, the attack can be demonstrated using other power-wasters discussed in prior literature, as they share similar structural configurations \cite{FPGAhammer} \cite{Krautter2019MitigatingCloud} \cite{9810438}. Importantly, FLARE is generalizable to other FPGAs as the vulnerabilities exploited during the partial reconfiguration process is not FPGA-specific. We use Vivado 2023.2 to synthesize, implement, and generate the bitstream corresponding to the hardware setup for the attack evaluation. Table \ref{lut} shows the resource utilization of the implemented designs. We use Vitis to load partial bitstreams into CoRQ.  We test the fault attack using three partial bitstreams -- \textit{blinkall, blinkcount}, and \textit{blinkline}. We monitor the attack success using the fault detection signal $flt_{sig}$, which is captured by the GPIO port of the FPGA. A video demo of the fault attack experiments performed in this work is available in \cite{demo}.

\vspace{-0.1cm}
 \begin{figure}
\centering
\includegraphics[width=0.43\textwidth]{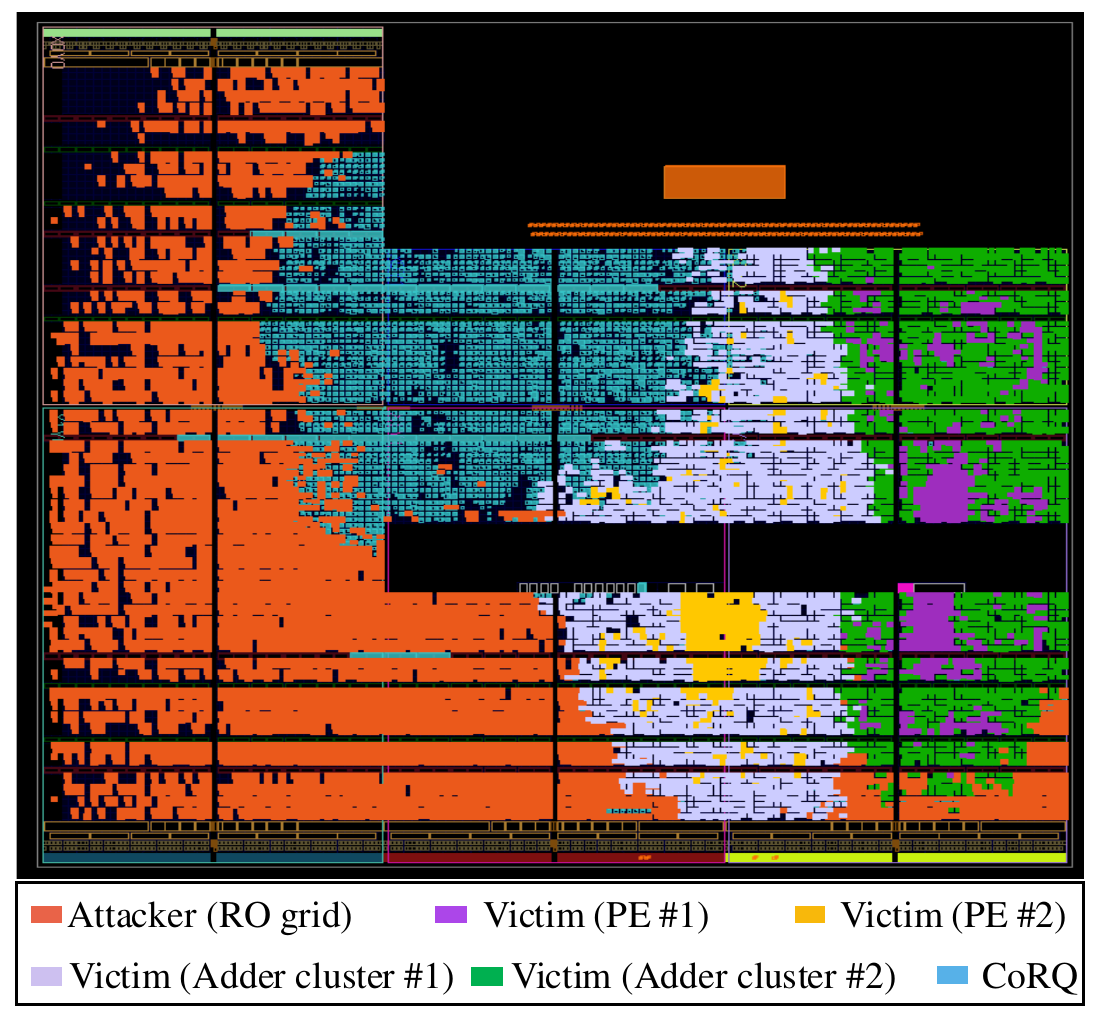}

\caption{Floorplan of the attack setup incorporating adder modules on Pynq-Z1 FPGA (PE: Priority encoder).}
\label{addfloorplan}
\vspace{-0.6cm}
\end{figure}
 \begin{figure}[t]

\includegraphics[width=0.5\textwidth]{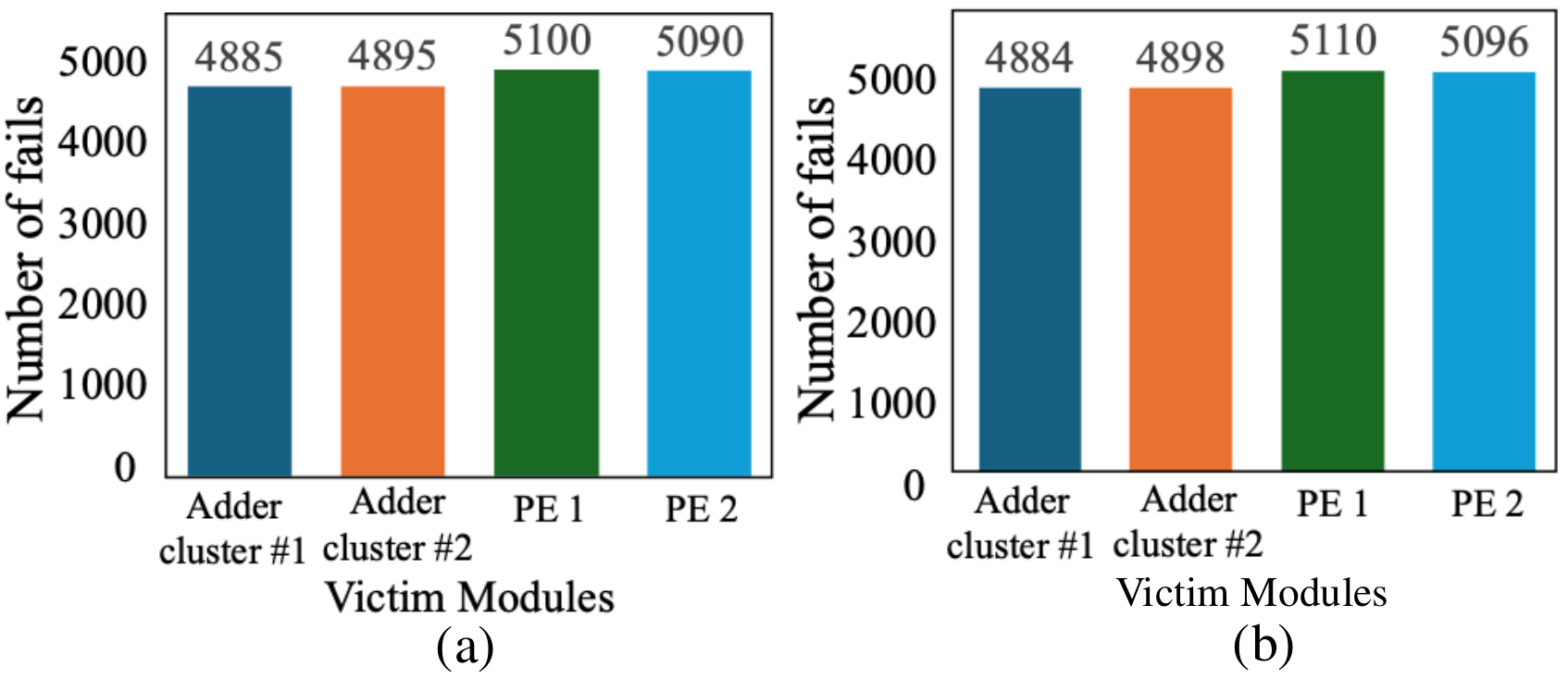}

\caption{Distribution of the number of fails in the adder modules for (a) RO and (b) Self-clocked RO (PE 1 and PE 2 refer to priority encoders $p_1$ and $p_2$, respectively).}
\label{adder}
\vspace{-0.5cm}
\end{figure}

\subsection{Fault Attack Analysis on Adders}
\begin{figure*}[t]

\includegraphics[width=\textwidth]{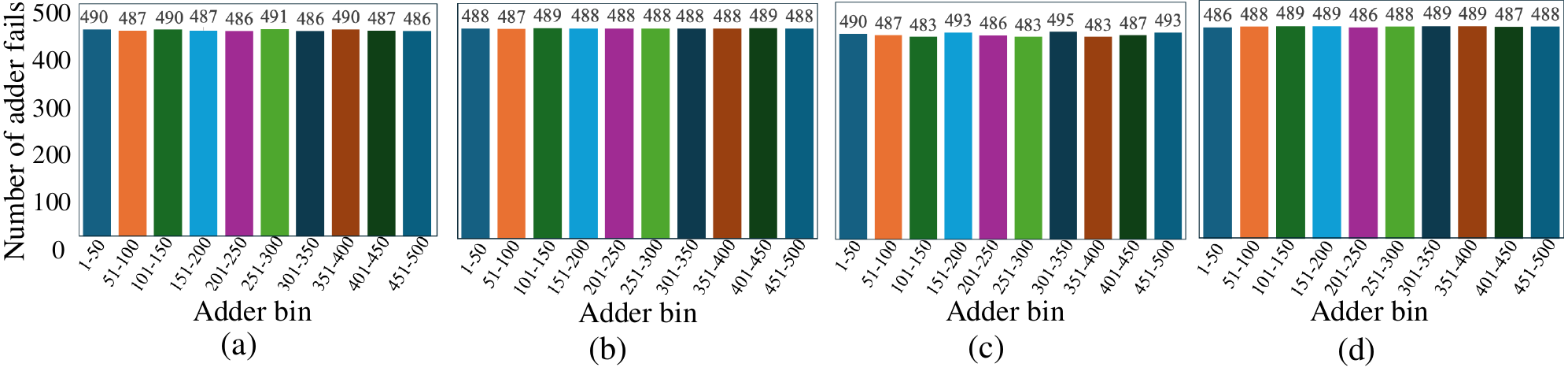}

\caption{ Parts (a) and (b) depict the number of fails in adder clusters \#1 and \#2, respectively, with ROs. Parts (c) and (d) show the fail frequency in clusters \#1 and \#2, respectively, with self-clocked ROs. }
\label{addf}
\vspace{-0.4cm}
\end{figure*}

\vspace{-0.1cm}
We evaluate FLARE on adder modules that are pre-configured on the FPGA. Fig. \ref{addfloorplan} shows the floorplan of the experimental setup implementing adder modules on the Pynq-Z1 FPGA. We conducted 10,000 fault attack attempts on the FPGA. In each attempt, a different partial bitstream is loaded into CoRQ for configuration. Power-wasters were activated specifically during the reconfiguration process to enable fault injection and were deactivated once the process concluded. After configuring the partial bitstream on the FPGA, we verify that the fault is successfully injected, altering the configuration address and redirecting the bitstream to the adder PRRs. Fig. \ref{adder} demonstrates the distribution of the fails due to the fault attack. We observe that for both combinational and self-clocked ROs, FLARE consistently induces faulty computations in at least one of the adder clusters$^1$. In some cases, the adders  remain unaffected, producing correct outputs. However, the faults disrupt the functionality of modules $p_1$ and $p_2$, resulting in incorrect encoded outputs. Fig. \ref{addf}(a) and Fig. \ref{addf}(b), show the frequency of fails in adder clusters \#1 and \#2, respectively. 



\subsection{Fault Attack Analysis on AES}
\begin{figure}
\centering
\includegraphics[width=0.43\textwidth]{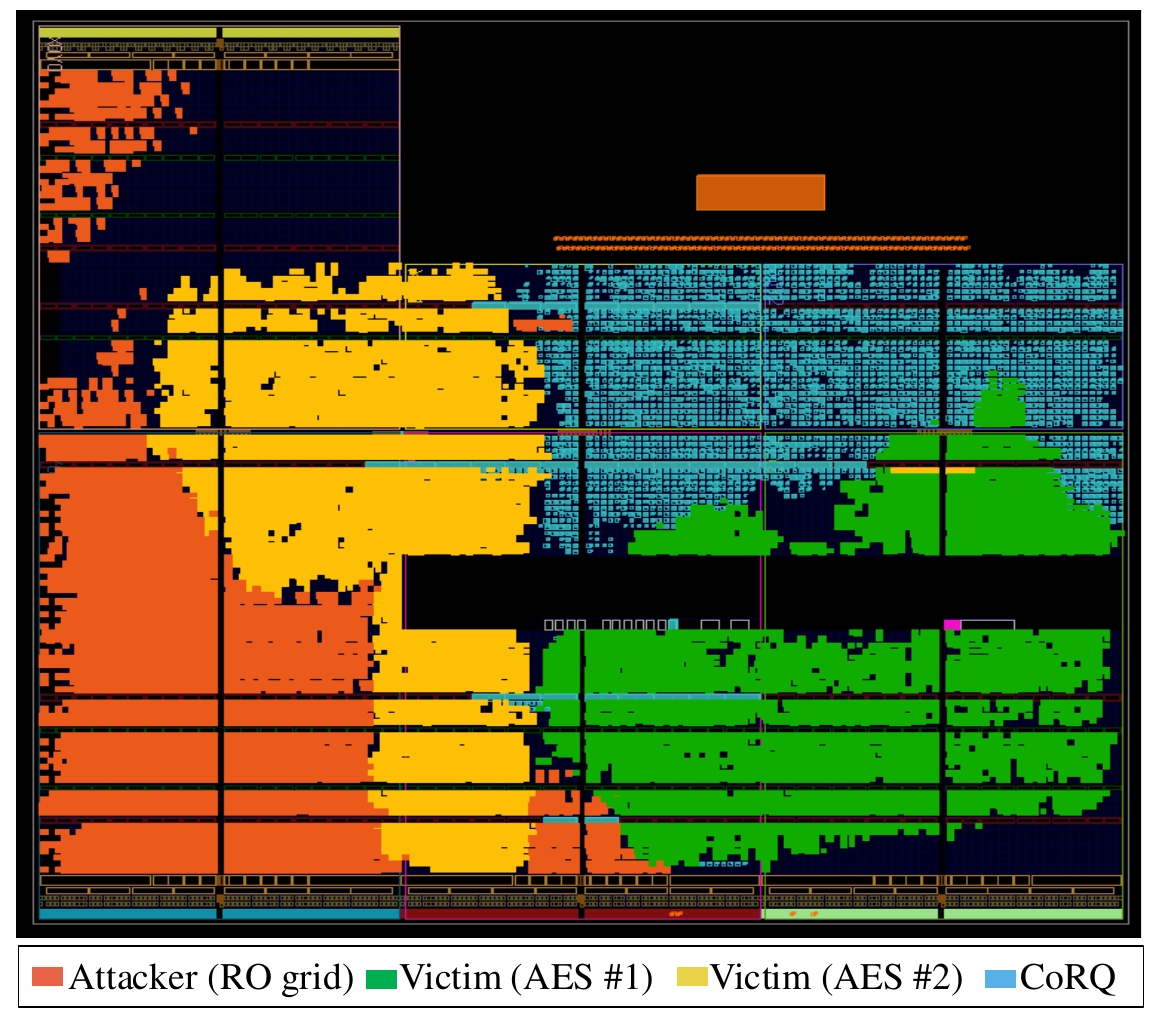}

\caption{Floorplan of the attack setup using AES modules.}
\label{aesfloorplan}
\vspace{-0.1cm}
\end{figure}

\vspace{-0.1cm}
To demonstrate the generalizability of FLARE, we evaluate another critical module, namely AES. The attack setup is shown in Fig. \ref{aesfloorplan}, which details the placement of the AES instances on the Pynq-Z1 FPGA. We keep the upper bound of AES instances to 2 as adding more instances exceeds the number of LUTs of the FPGA, causing place-and-route errors during the Vivado implementation phase. Similar to adder evaluation, we performed a total of 10,000 fault attack attempts using both combinational and self-clocked ROs. We observe successful fault-injection by both the power-wasters --  after five fault-injection attempts, AES \#1 consistently generates incorrect ciphertexts, indicating successful disruption of its encryption operations\footnote[1]{A video demo of the fault attack is presented in \cite{demo}.}. However, AES \#2 remains unaffected by the fault attack, primarily due to the placement and routing constraints of the PRRs where it is configured.

\vspace{-0.1cm}
\subsection{Bypassing CRC Mechanisms}
\vspace{-0.1cm}
During bitstream generation, the CRC value is computed and embedded within the bitstream. The \textit{crc\_check} function in CoRQ computes the CRC of the bitstream before configuring it on the FPGA. As explained in \cite{crc}, CRC only verifies the bitstream data; there are no checks to ensure that the bitstream is configured in the correct PRR. 
  To validate whether FLARE evades CRC, we conducted experiments using a test bitstream, namely \textit{blinkcount}. We monitor the \textit{corq\_status} flag to check if CRC detects a faulty bitstream. We obtain the following message on Vitis: \textit{Decrypting blinkcount successful. Bitstream successfully loaded.} Since FLARE manipulates the configuration address rather than the bitstream data, it successfully evades CRC, and the fault-injected bitstream is incorrectly configured in the victim PRR of the FPGA.
\vspace{-0.1cm}

\subsection{Comparison to Prior Fault Attack Schemes}
\vspace{-0.1cm}

By precisely targeting the configuration address of a bitstream and reducing the attack duration, FLARE is significantly faster and stealthier than prior runtime fault attacks \cite{8056840} \cite{FPGAhammer} \cite{7809042} \cite{8844478}, while minimizing the likelihood of detection by \cite{9643485} \cite{10.1145/3451236}. Moreover, FLARE adversely impacts multiple modules on the FPGA simultaneously, a threat not explored in prior fault attacks. For a fair evaluation, we compare FLARE with  \cite{chaudhuri2024hackingfabrictargetingpartial}, as both approaches specifically target fault-injection in the partial reconfiguration process. As highlighted in Table \ref{faultcomp}, FLARE is 200$\times$ faster than \cite{chaudhuri2024hackingfabrictargetingpartial} in executing a successful attack. While both \cite{chaudhuri2024hackingfabrictargetingpartial} and FLARE successfully induce faults in the bitstream, \cite{chaudhuri2024hackingfabrictargetingpartial} has a 100\% detection rate by CRC, whereas FLARE bypasses CRC verification entirely.  

\vspace{-0.1cm}
\begin{table}[t]
\centering
\fontsize{8.7}{8.7}\selectfont

\caption{Comparison of FLARE with \cite{chaudhuri2024hackingfabrictargetingpartial}.}

    \label{compare}
    \begin{tabular}{|c| c| c|} 
    \hline
{Parameter} &\cite{chaudhuri2024hackingfabrictargetingpartial} & \textbf{FLARE} \\
\hline
Attack duration&40 ms&200 $\mu$s \\
\hline
Exposure window&Entire bitstream &Targeted (`select' part \\
&&of bitstream) \\
\hline
CRC evasion?&\xmark&\checkmark \\
\hline
\end{tabular}

\vspace{-0.3cm}

\label{faultcomp}
\end{table}

\section{Conclusion}
\vspace{-0.1cm}
We have presented FLARE, a novel fault attack that specifically targets the process of partial reconfiguration of a bitstream, and manipulates the configuration address of the bitstream. Experiments on several design modules demonstrate the effectiveness of FLARE in causing faulty computation by these modules as well as DoS.  Simultaneously impacting multiple tenants while reducing the chance of detection highlights the severity of FLARE in multi-tenant FPGA environments.

\bibliographystyle{IEEEtran}
{
\hyphenpenalty=10000
\exhyphenpenalty=10000
\sloppy
\bibliography{references}
\vspace{-0.4cm}
}

\end{document}